\documentclass{aa} % for a referee version
\pdfoutput=1
\usepackage{graphicx}
\usepackage{txfonts}
\usepackage[utf8]{inputenc}
\usepackage[table]{xcolor}
\usepackage{array}
\usepackage{float}
\usepackage[]{natbib}
\usepackage[colorlinks=true,linkcolor=blue,citecolor=blue]{hyperref}
\begin{document} 

    \title{The LEECH Exoplanet Imaging Survey. Further constraints on the planet architecture of the HR~8799 system\thanks{The LBT is an international collaboration among institutions in the United States, Italy and Germany. LBT Corporation partners are: The University of Arizona on behalf of the Arizona university system; Istituto Nazionale di Astrofisica, Italy; LBT Beteiligungsgesellschaft, Germany, representing the Max-Planck Society, the Astrophysical Institute Potsdam, and Heidelberg University; The Ohio State University, and The Research Corporation, on behalf of The University of Notre Dame, University of Minnesota, and University of Virginia.}}

   \author{A.-L.~Maire\inst{1}, A.\,J.~Skemer\inst{2,}\thanks{Hubble Fellow}, P.\,M.~Hinz\inst{2}, S.~Desidera\inst{1}, S.~Esposito\inst{3}, R.~Gratton\inst{1}, F. Marzari\inst{4}, M.\,F.~Skrutskie\inst{5}, B.\,A.~Biller\inst{6,7}, D.~Defr\`ere\inst{2}, V.\,P.~Bailey\inst{2}, J.\,M.~Leisenring\inst{2}, D.~Apai\inst{2}, M.~Bonnefoy\inst{8,9,7}, W.~Brandner\inst{7}, E.~Buenzli\inst{7}, R.\,U.~Claudi\inst{1}, L.\,M.~Close\inst{2}, J.\,R.~Crepp\inst{10}, R.\,J.~De Rosa\inst{11,12}, J.\,A.~Eisner\inst{2}, J.\,J.~Fortney\inst{13}, T.~Henning\inst{7}, K.-H.~Hofmann\inst{14}, T.\,G.~Kopytova\inst{7,15}, J.\,R.~Males\inst{2,}\thanks{NASA Sagan Fellow}, D.~Mesa\inst{1}, K.\,M.~Morzinski\inst{2,}\thanks{NASA Sagan Fellow}, A.~Oza\inst{5}, J.~Patience\inst{11}, E.~Pinna\inst{3}, A.~Rajan\inst{11}, D.~Schertl\inst{14}, J.\,E.~Schlieder\inst{7,16,}\thanks{NASA Postdoctoral Program Fellow}, K.\,Y.\,L.~Su\inst{2}, A.~Vaz\inst{2}, K.~Ward-Duong\inst{11}, G.~Weigelt\inst{14}, and C.\,E.~Woodward\inst{17}}

   \institute{INAF -- Osservatorio Astronomico di Padova, Vicolo dell'Osservatorio 5, 35122 Padova, Italy\\
              \email{annelise.maire@oapd.inaf.it}              
			  \and                            
              Steward Observatory, Department of Astronomy, University of Arizona, 993 North Cherry Avenue, Tucson, AZ~85721, USA              
              \and              
              INAF -- Osservatorio Astrofisico di Arcetri, Largo E. Fermi 5, 50125 Firenze, Italy
              \and
              Dipartimento di Fisica e Astronomia, Universit\'a di Padova, via F. Marzolo 8, 35131 Padova, Italy
              \and
              Department of Astronomy, University of Virginia, Charlottesville, VA~22904, USA
              \and
              Institute for Astronomy, University of Edinburgh, Blackford Hill, Edinburgh EH9 3HJ, United Kingdom
              \and
              Max-Planck-Institut f\"ur Astronomie, K\"onigstuhl 17, 69117 Heidelberg, Germany
              \and
              Universit\'e Grenoble Alpes, IPAG, 38000 Grenoble, France
              \and
              CNRS, IPAG, F-38000 Grenoble, France
              \and
              Department of Physics, University of Notre Dame, 225 Nieuwland Science Hall, Notre Dame, IN 46556, USA
              \and
              Arizona State University, School of Earth and Space Exploration, PO Box 871404, Tempe, AZ~85287-1404, USA
              \and
              Astrophysics group, School of Physics, University of Exeter, Stocker Road, Exeter EX4 4QL, United Kingdom
              \and
              Department of Astronomy and Astrophysics, University of California Santa Cruz, Santa Cruz, CA~95064, USA
              \and
              Max-Planck-Institut f\"ur Radioastronomie, Auf dem H\"ugel 69, 53121 Bonn, Germany
              \and
              International Max Planck Research School for Astronomy and Space Physics, Heidelberg, Germany
              \and
              NASA Ames Research Center, M.S. 245-6, Moffett Field, CA~94035, USA
              \and
              Minnesota Institute for Astrophysics, University of Minnesota, 116 Church Street, SE, Minneapolis, MN 55455, USA
              }

   \date{Received 20 October 2014 / Accepted 4 March 2015}

% \abstract{}{}{}{}{} 
% 5 {} token are mandatory
 
  \abstract
  % context heading (optional)
  % {} leave it empty if necessary  
   {Astrometric monitoring of directly-imaged exoplanets allows the study of their orbital parameters and system architectures. Because most directly-imaged planets have long orbital periods ($>$20~AU), accurate astrometry is challenging when based on data acquired on timescales of a few years and usually with different instruments. The LMIRCam camera on the Large Binocular Telescope is being used for the LBT Exozodi Exoplanet Common Hunt (LEECH) survey to search for and characterize young and adolescent exoplanets in $L^{\prime}$ band (3.8~$\muup$m), including their system architectures.}
  % aims heading (mandatory)
   {We first aim to provide a good astrometric calibration of LMIRCam. Then, we derive new astrometry, {test the predictions of the orbital model of 8:4:2:1 mean motion resonance proposed by Go\'zdziewski \& Migaszewski, and perform new orbital fitting of the HR~8799~bcde planets.} We also present {deep limits on a putative fifth planet} interior to the known planets.}
  % methods heading (mandatory)
   {We use observations of HR~8799 and the $\Theta^1$ Ori C field obtained during the same run in October 2013.}
  % results heading (mandatory)
   {We first characterize the distortion of LMIRCam. We determine a platescale and a {true north} orientation for the images of {10.707\,$\pm$\,0.012~mas/pix and $-$0.430\,$\pm$\,0.076$^{\circ}$}, respectively. The errors on the platescale and {true north} orientation translate into astrometric accuracies at a separation of 1$''$ of 1.1~mas and 1.3~mas, respectively. {The measurements for all planets are in agreement within 3~$\sigma$ with the ephemeris predicted by Go\'zdziewski \& Migaszewski. The orbital fitting based on the new astrometric measurements favors an architecture for the planetary system based on {8:4:2:1} mean motion resonance.} The detection limits allow us to exclude a fifth planet slightly brighter/more massive than HR~8799 b at the location of the {2:1} resonance with HR~8799 e ($\sim$9.5~AU) and about twice as bright as HR~8799 cde at the location of the {3:1} resonance with HR~8799 e ($\sim$7.5~AU).}
   %conclusions (optional)
   {}


\authorrunning{A.-L. Maire et al.}
\titlerunning{The LEECH Exoplanet Imaging Survey. Further constraints on the HR~8799 planet architecture}

   \maketitle
%   _______________________________________________________________

\section{Introduction}

The detection of exoplanets and the characterization of their atmospheres and system architectures using direct imaging is one of the most difficult challenges in modern astronomy. Optimized observing strategy and data analysis are required to overcome the high contrasts ($\gtrsim$10$^4$) and the small separations (a few {tenths} of an arcsecond) between a star and a planet. The development of adaptive optics systems, coronagraphic devices, and differential imaging techniques in the {past fifteen years} allowed the detection of planetary-mass objects in favorable situations: young and nearby host star, large orbital separation, and/or low star/planet mass ratio \citep[e.g,][]{Chauvin2005a, Marois2008c, Marois2010b, Lagrange2010b, Lafreniere2010, Kuzuhara2013, Rameau2013b, Bailey2014}. A new generation of instruments dedicated to the search and characterization of young exoplanets down to the Jupiter mass {has started operations} {\citep{Tamura2010, Hinkley2011, Close2013, Skemer2014c, Macintosh2014, Beuzit2012}}.

One main topic of the study of exoplanetary systems using direct imaging is their architecture. {This can be investigated using indirect or direct observational evidence of the presence of planets}. The modeling of the dust spatial distribution in resolved circumstellar debris disks can provide predictions on the orbital parameters and the mass of one or more putative planets which gravitationnally perturb the disk \citep[e.g,][]{Mouillet1997, Wyatt1999, Ozernoy2000, Augereau2001, Kalas2005}. These predictions may be followed by the {direct detection of a planet \citep{Lagrange2009, Lagrange2010b} or a substellar-mass object {\citep{Kalas2008, Currie2012, Galicher2013, Kalas2013}}.} When an object is detected, {accurate astrometric follow-up is needed} to determine its orbital elements \citep{Chauvin2012, Kalas2013}. As most of the directly-imaged exoplanets have long orbital periods ($>$20~AU), this analysis is challenging when based on data acquired on timescales of a few years and usually with different instruments. {For multiple-planet systems, the study of the dynamical stability allows to put constraints on the planet masses \citep{Fabrycky2010, Gozdziewski2009, Reidemeister2009, MoroMartin2010, Marois2010b, Currie2011, Sudol2012, Esposito2013, Gozdziewski2014}. These estimates are not dependent on evolutionary models of giant planets \citep[e.g.,][]{Burrows1997, Chabrier2000, Baraffe2003, Marley2007}, which rely on unknown initial conditions \citep{Marley2007, Spiegel2012, Marleau2014} and poorly constrained stellar ages. Thus,} the derivation of the dynamical mass of young low-mass companions may help to calibrate the evolutionary models {\citep[e.g.,][]{Close2005, Boden2005, Crepp2012, Bonnefoy2014c}}.

\begin{table*}[t]
\caption{Log of the observations.}
\label{tab:obs}
\begin{center}
\begin{tabular}{l c c c c c c c c c}
\hline\hline
Object & Obs. date (UT) & Seeing ($''$) & Air mass start/end & DIT (s) & NDIT & $N_{\mathrm{exp}}$ & $N_{\mathrm{dith}}$ & $\Delta$PA ($^{\circ}$) & Remarks \\
\hline
$\Theta^1$ Ori C & 2013/10/24 & 0.95--1.00 & 1.50--1.47 & 0.029 & 30 & 10 & 12 & 1.728 & Unsaturated \\
HR~8799 & 2013/10/21 & 0.70--1.10 & 1.05--1.15 & 0.291 & 3 & 7223 & -- & 107 & Saturated \\
\hline
\end{tabular}
\end{center}
\tablefoot{The seeing is the value measured by the LBT differential image motion monitor (DIMM) in the same direction on sky as the target. DIT {(Detector Integration Time)} refers to the exposure time per coadd, NDIT {(Number of Detector InTegrations)} to the number of coadds for a single frame, $N_{\rm{exp}}$ to the number of frames per dither pattern, $N_{\rm{dith}}$ to the number of dither patterns, and $\Delta$PA to the amplitude of the parallactic rotation.} \\
\end{table*}

The LBT Exozodi Exoplanet Common Hunt \citep[LEECH,][]{Skemer2014c} survey to search for and characterize young and adolescent exoplanets in $L^{\prime}$ band (3.8~$\muup$m) started at the Large Binocular Telescope in February 2013. This {$\sim$130}-night survey exploits {the adaptive optics system FLAO} \citep{Esposito2010}, which combines Strehl ratios performance superior to 80\% in $H$ band and 95\% in $L^{\prime}$ band with low thermal emissivity, the LBT Interferometer \citep[LBTI,][]{Hinz2008}, and the $L$/$M$-band InfraRed Camera \citep[LMIRCam,][]{Skrutskie2010, Leisenring2012}. The selection of the $L^{\prime}$ band is strategic, because AO systems provide better Strehl ratios and the star/planet brightness ratios are reduced for this spectral band with respect to {shorter wavelengths.} LEECH complements the other current or future large imaging surveys for young ($\lesssim$200~Myr) giant exoplanets by covering longer wavelengths, and probing northern, nearby ($\lesssim$30--55~pc), and older ($\lesssim$1~Gyr) {stars. The first data obtained} with LBTI/LMIRCam prior to the LEECH survey allowed the multi-wavelength photometric analysis in the $L$ band (3--4~$\muup$m) of the four planets of HR~8799 \citep{Skemer2012, Skemer2014b}, and the $M$-band detection and photometric characterization of $\kappa$~And~b \citep{Bonnefoy2014a}.

One important aspect of {a high-contrast imaging survey is the accurate determination of the astrometry of the detected exoplanet candidates. To date,} only one astrometric analysis based on LBTI/LMIRCam data and {not related to the LEECH primary goals} has been performed, for the white dwarf companion to HD~114174 \citep{CMatthews2014}, using observations of an astrometric binary but without correcting for the distortion effects of the camera.

We present in this paper the first reliable LEECH astrometry of an exoplanetary system, HR~8799. {The host star is a young \citep[30~Myr,][]{Marois2010b, Baines2012} and nearby \citep[$d$\,=\,39.4\,$\pm$\,1.0~pc,][]{VanLeeuwen2007} late-A or early-F star \citep{Gray1999, Gray2014}. It is orbited by at least four giant planets at projected separations of 15, 24, 38, and 68~AU \citep{Marois2008c, Marois2010b}. The system is also composed of a complex debris disk \citep{Su2009} seen with an inclination of 26$^{\circ}$ \citep{BMatthews2014}, with a warm ($T$\,$\sim$\,150~K) and unresolved inner dust component between about 6 and 15~AU, a cold ($T$\,$\sim$\,35~K) planetesimal belt extending from $\sim$100 to 310~AU, and a cold halo of dust grains which extends from 310 to $\sim$2000~AU.} We also describe in this paper the astrometric calibration and characterization of the LMIRCam distortion, which are of general interest for the users of the instrument (now offered to the LBT community). We also present new orbital fitting of all four planets and further constraints on the properties of a putative fifth planet interior to these planets. We describe the observations and the data reduction in Sect.~\ref{sec:observations} and Appendix~\ref{sec:appendixdistortion}. We present the astrometric analysis in Sect.~\ref{sec:astrometry}. We report the planet photometry and the detection limits in Sects.~\ref{sec:photom} and \ref{sec:detlims}. We perform new orbital fitting of the astrometric measurements in Sect.~\ref{sec:orbitarchitect}. We discuss our results and assumptions in Sect.~\ref{sec:discussion}.

\section{Observations and data reduction}
\label{sec:observations}
The observations of HR~8799 and the $\Theta^1$~Ori~C field were carried out during the same observing run in October 2013 (Table~\ref{tab:obs}). The LBTI is located at a bent Gregorian focus with no field rotator, hence all observations are performed in pupil-stabilized mode. This mode allows the use of angular differential imaging \citep{Marois2006a} to subtract the point-spread function (PSF) for high-contrast imaging. {The LMIRCam camera is designed to be Nyquist-sampled for the full 22.8-m aperture of the LBT when used in interferometric mode. When LMIRCam is used with one of the two apertures of 8.4~m, the images are oversampled. For the high-contrast imaging observations (HR~8799 in this paper), we bin the pixels 2 by 2.}

The $\Theta^1$~Ori~C field was observed on October 24th UT {with the right side (also known as the ``DX'' side) of the LBT} at twelve different dither patterns, with each dither position consisting of ten frames. For each dither position, the frames are corrected for the distortion effects of the camera (Appendix~\ref{sec:appendixdistortion}) and the bad pixels, sky-subtracted, and averaged. The mean measured PSF FWHM is {$\sim$10~pixels estimated by several methods} (gaussian fitting, radial profile).

HR~8799 \citep[$L^{\prime}$\,=\,5.220\,$\pm$\,0.018,][]{Hinz2010} was observed on October 21st UT with the ``DX'' side of the telescope at two different nods {simultaneously}. {We selected the individual exposure time so that the readout noise is kept below the sky background. The star is thus saturated on the detector.} After applying frame selection, distortion correction, cosmetic removal, sky subtraction, 2$\times$2 binning, {and frame registration (using cross-correlation),} the data are processed using our implementation of the principal component analysis approach \citep[PCA,][]{Soummer2012, Amara2012}. Figure~\ref{fig:hr8799image} shows the resulting image. Planets bcde are detected with signal-to-noise ratio {81, 154, 62, and 32 (see below for the method used for the noise estimation). Immediately after observing HR 8799, we observed another star, HIP 18859, in the same mode as HR 8799 but with a calibrated 1\% neutral density filter. When observing a bright star in the $L^{\prime}$ band, the LMIRCam PSF is very stable, so this photometric standard is more than adequate to approximate the relatively low signal-to-noise core of an exoplanet. Since the point-source self-subtraction} of angular differential imaging biases the astrometric and photometric measurements, we calibrate these effects using injection of synthetic point sources (based on the measured PSF) with different brightness and insertion positions {into the pre-processed data} \citep{Skemer2012}. More precisely, we insert negative artificial planets {\citep{Marois2010a, Bonnefoy2011}} at the positions of the real planets {in the pre-processed data before performing the differential imaging part of the data analysis}. This is repeated as part of a Levenberg-Marquardt fit \citep{Markwardt2009}\footnote{\label{note:mpfit} \url{http://purl.com/net/mpfit}.} until the flux at the positions of the true planets is minimized. {The procedure minimizes the planet flux over an annulus centered on the star with a radius equal to the planet separation and a width of 0.3$''$. {Using an annulus for the minimization zone instead of an aperture around the planet PSF has negligible effects on the results of the minimization process because a planet signal spans a small region of the annulus. The Levenberg-Marquardt code outputs error bars, based on the cross-correlation matrix and the number of free parameters (i.e. the number of resolution elements) in the annulus.}

   \begin{figure}[t]
   \centering
   \includegraphics[trim = 35mm 35mm 35mm 35mm, clip,width=.45\textwidth]{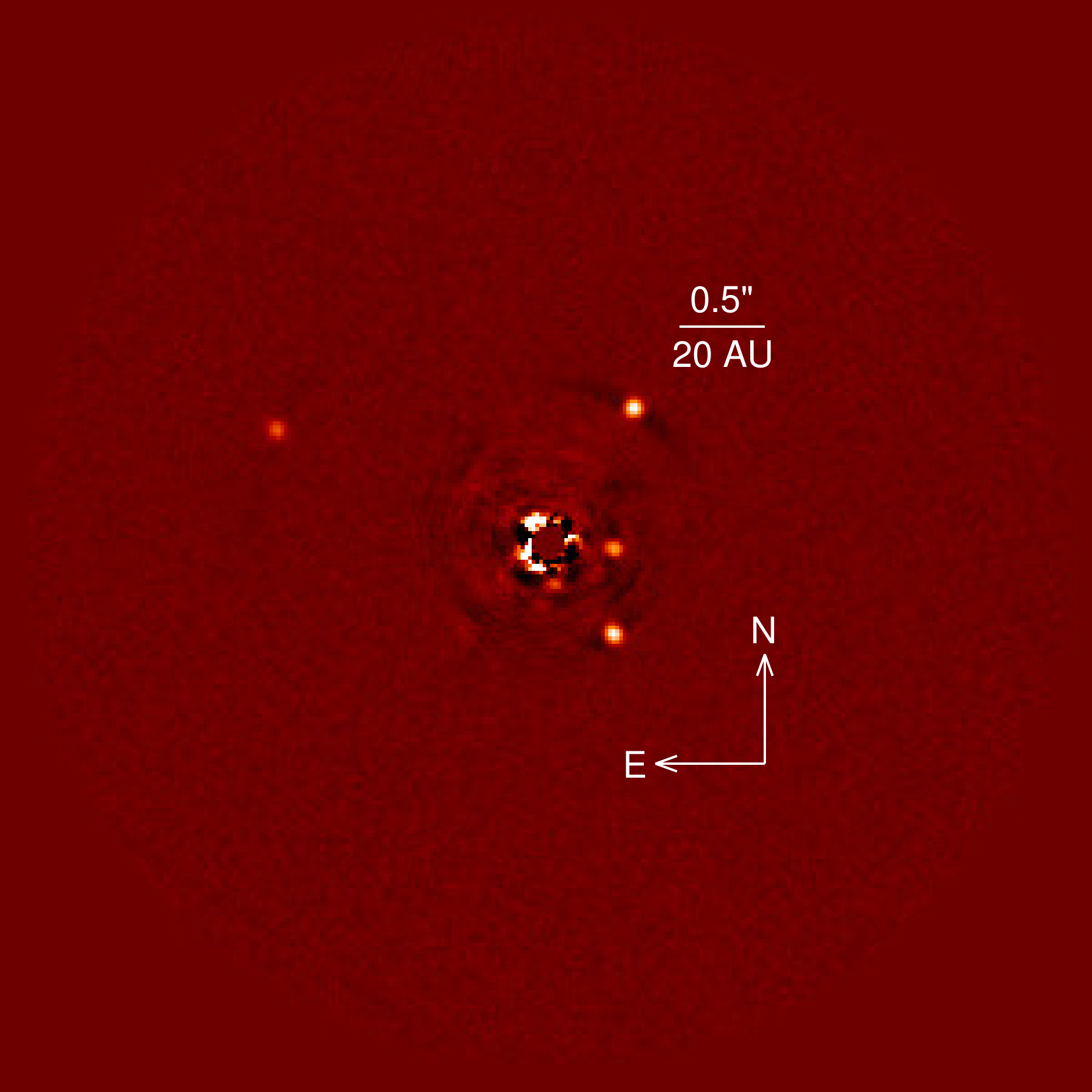}
      \caption{LMIRCam image in $L^{\prime}$ band of the HR~8799 multiple-planet system. The field of view is $\sim$4$''$. The point-spread function is subtracted using our PCA pipeline (see text). The image is binned (2$\times$2 binning). The intensity scale is linear.
              }
         \label{fig:hr8799image}
   \end{figure}

We construct a contrast curve (discussed in Sect.~\ref{sec:detlims}) using the pre-processed frames with HR~8799 bcde artificially removed. At every radius, we insert a fake planet in these data and perform a complete re-reduction, iterating until we reach a 5-$\sigma$ detection\footnote{{As recently discussed in \citet{Mawet2014}, the confidence level associated to a given $\sigma$ threshold decreases when the separation decreases, due to the small sample statistics of the noise realizations. For the separation range considered in this work ($\gtrsim$2~$\lambda/D$), it is superior to the confidence level corresponding to a 3-$\sigma$ gaussian threshold.}} (our nominal contrast benchmark). {For each separation to the star, the noise level is estimated using the standard deviation of the intensity of the pixels at this separation. For the derivation of the signal-to-noise of the planets, the noise level is scaled to the same aperture size (1~$\lambda/D$) used to estimate the flux of the fake planets assuming white noise. We repeat this analysis} at eight position angles for every radius, and average the results to suppress random speckle error.

\section{Astrometric analysis}
\label{sec:astrometry}

\subsection{The $\Theta^1$~Ori~C field}
{The stellar positions in the images are derived using the centroid IDL routine \texttt{cntrd}\footnote{\url{http://idlastro.gsfc.nasa.gov/ftp/pro/idlphot/}.} adapted from the DAOphot software \citep{Stetson1987}. For the catalog positions, we use the astrometry published in \citet{Close2012}, based on LBT/PISCES observations acquired on October 16th, 2011, which is referenced to the HST/ACS observations of \citet{Ricci2008}. In each of the dithered frames of the observing sequence (Sect.~\ref{sec:observations}), we select typically 3--4 stars with signal-to-noise ratios greater than 10 for the calibration. We use all the available stellar pairs for the analysis.}

\begin{table*}[t]
\caption{Astrometric measurements of the HR~8799 planets with respect to the star at epoch 2013.81.}
\label{tab:astrometry}
\begin{center}
\begin{tabular}{l c c c c}
\hline\hline
Planet & $\Delta$RA ($''$) & $\Delta$Dec ($''$) & {Separation} ($''$) & {Parallactic angle} ($^{\circ}$)\\
\hline
b & 1.5624\,$\pm$\,0.0085 & 0.7133\,$\pm$\,0.0130 & 1.7176\,$\pm$\,0.0131 & 65.46\,$\pm$\,0.44\\
c & $-$0.5383\,$\pm$\,0.0060 & 0.7838\,$\pm$\,0.0131 & 0.9508\,$\pm$\,0.0142 & {325.52\,$\pm$\,0.87} \\
d & $-$0.3771\,$\pm$\,0.0070 & $-$0.5380\,$\pm$\,0.0111 & 0.6571\,$\pm$\,0.0134 & 215.0\,$\pm$\,1.2 \\ 
e & $-$0.3938\,$\pm$\,0.0105 & $-$0.0357\,$\pm$\,0.0168 & 0.3954\,$\pm$\,0.0119 & 264.8\,$\pm$\,1.7 \\
\hline
\end{tabular}
\end{center}
\tablefoot{{The values of $\Delta$RA and $\Delta$Dec are the averages of two measurements (see text). The error for $\Delta$RA is the average of the statistical errors associated to the individual measurements. The error for $\Delta$Dec is estimated using the quadratic combination of the statistical errors associated to the individual measurements and of the bias between the individual measurements. The statistical error of an individual measurement is derived using the quadratic combination of the error terms detailed in Table~\ref{tab:errorbudget}.}} \\
\end{table*}

{Combining the results obtained for the whole dataset, we derive forty individual measurements for the platescale and the {true north} orientation with respect to the detector y-axis (counted positively in the counterclockwise direction). The mean value and the corresponding error are:
\begin{itemize}
\item Plate scale: 10.707\,$\pm$\,0.012~mas/pix
\item True north: $-$0.430\,$\pm$\,0.076$^{\circ}$
\end{itemize}
The errors on the platescale and {true north} orientation translate into astrometric accuracies at a separation of 1$''$ of 1.1~mas and 1.3~mas, respectively.}

{We do not correct the catalog positions for the differential stellar proper motions \citep[the absolute motions are expected to be $\sim$1.5~mas/yr,][]{Close2012}. The error on the platescale produced by the differential stellar proper motions is negligible for the separations of the stellar pairs considered in this work (mean value 5$''$), with respect to the errors induced by the platescale estimation (relative error 0.1\%, Table~\ref{tab:errorbudget}) and the detector distortion (relative error $\sim$0.1\% after correction, see Appendix~\ref{sec:appendixdistortion}).}

\begin{table}[t]
\caption{1-$\sigma$ astrometric error budget (in mas) for the HR~8799 planets for both observations (see text).}
\label{tab:errorbudget}
\begin{center}
\begin{tabular}{l c c c c c}
\hline\hline
Error source & Error & b & c & d & e \\
\hline
Star center & 0.25 pix & 5 & 5 & 5 & 5 \\
Platescale & 0.1\% & 2 & 1 & 1 & 1 \\
{True north} & 0.076$^{\circ}$ & 2 & 1 & 1 & 1 \\
Distortion & 0.1\% & 2 & 1 & 1 & 1 \\
Fitting & & 5/6 & 3/3 & 5/4 & 11/7 \\
\hline
\end{tabular}
\end{center}
\tablefoot{The error sources are the determination of the star center, the camera platescale, the orientation of the {true north} after derotation of the images, the solution used for the modeling of the distortion effects, and the photometric biases induced by the PCA. We give the fitting error in both horizontal and vertical directions for each observation (Sect.~\ref{sec:observations}).} \\
\end{table}

\subsection{The HR~8799 planetary system}

{HR~8799 was observed once during the night of October 21st 2013 simultaneously at two different nods. {We combine the individual measurements of these two datasets for each of the four planets to derive the astrometric measurements in Table~\ref{tab:astrometry}. We give the error budget for both datasets in Table~\ref{tab:errorbudget}. We note a bias of $\sim$0.015$''$ between the two sets of measurements for the relative vertical position. This may be due to a less accurate centering of the frames (Sect.~\ref{sec:observations}). We account for this bias in the derivation of the error on the relative vertical position given in Table~\ref{tab:astrometry}.} The star center and the fitting error are the dominant error sources. For planets b and d, both sources have similar contributions. For planet c, the star center is the main error source. For planet e, the fitting error is the largest error. {The accuracies on our measurements are similar to those measured by \citet{Currie2014} in Keck $L^{\prime}$ data for the three outer planets. For the innermost planet, the accuracy is worse with respect to the work of \citet{Currie2014} because of the large fitting error, but similar to the values measured in Keck $L^{\prime}$ data in the discovery paper \citep{Marois2010b}.}

\begin{table}[t]
\caption{{LBTI/LMIRCam relative and absolute photometry of the HR~8799 planets in $L^{\prime}$ band.}}
\label{tab:photometry}
\begin{center}
\begin{tabular}{l c c}
\hline\hline
Planet & $\Delta L^{\prime}$ with HR~8799 c & $M_{L^{\prime}}$ \\
\hline
b & $+$0.92\,$\pm$\,0.08 & 12.66\,$\pm$\,0.12 \\
c &  & 11.74\,$\pm$\,0.09 \\
d & $-$0.07\,$\pm$\,0.11 & 11.67\,$\pm$\,0.14 \\
e & $+$0.13\,$\pm$\,0.13 & 11.87\,$\pm$\,0.16 \\
\hline
\end{tabular}
\end{center}
\tablefoot{{HR~8799 c absolute photometry from \citet{Marois2008c}. The relative photometry for the other planets is derived using two datasets (Sect.~\ref{sec:observations}). For HR~8799 be, there is no clear bias between the two individual measurements and the error on the relative photometry is the average of the statistical errors of the individual measurements. For HR~8799 d, the error on the relative photometry is the quadratic combination of the statistical errors and of the bias ($\sim$0.08~mag). The statistical error on the individual measurements includes the fitting error and the stellar flux variations during the observation (1.1\%, see text). The error bar on the absolute photometric values includes the uncertainty on the HR~8799 c absolute photometry.}} \\
\end{table}

\section{Photometry}
\label{sec:photom}

   \begin{figure*}[t]
   \centering
   \includegraphics[trim = 13mm 5mm 10mm 10mm, clip,width=.47\textwidth]{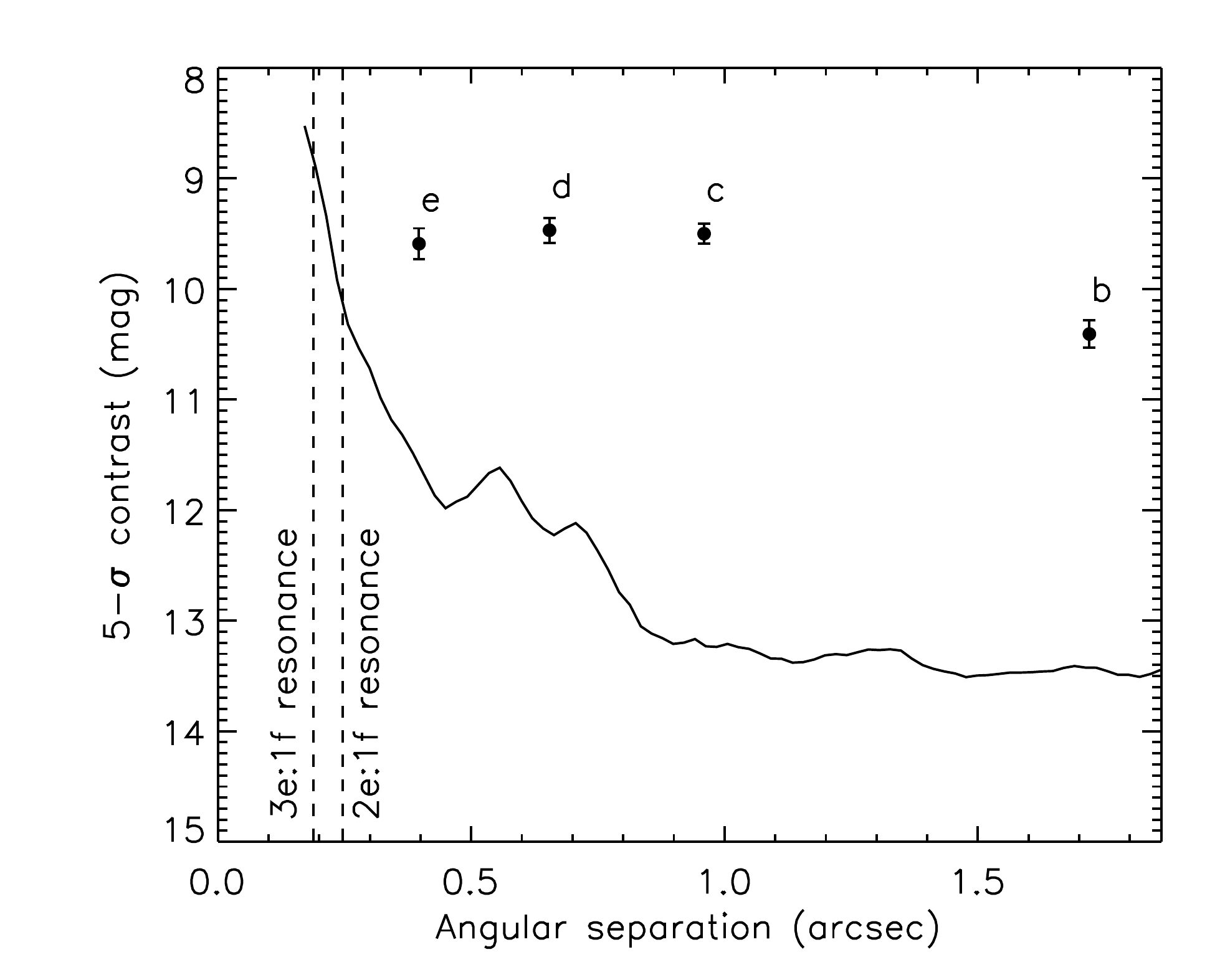}
   \includegraphics[trim = 13mm 5mm 10mm 10mm, clip,width=.47\textwidth]{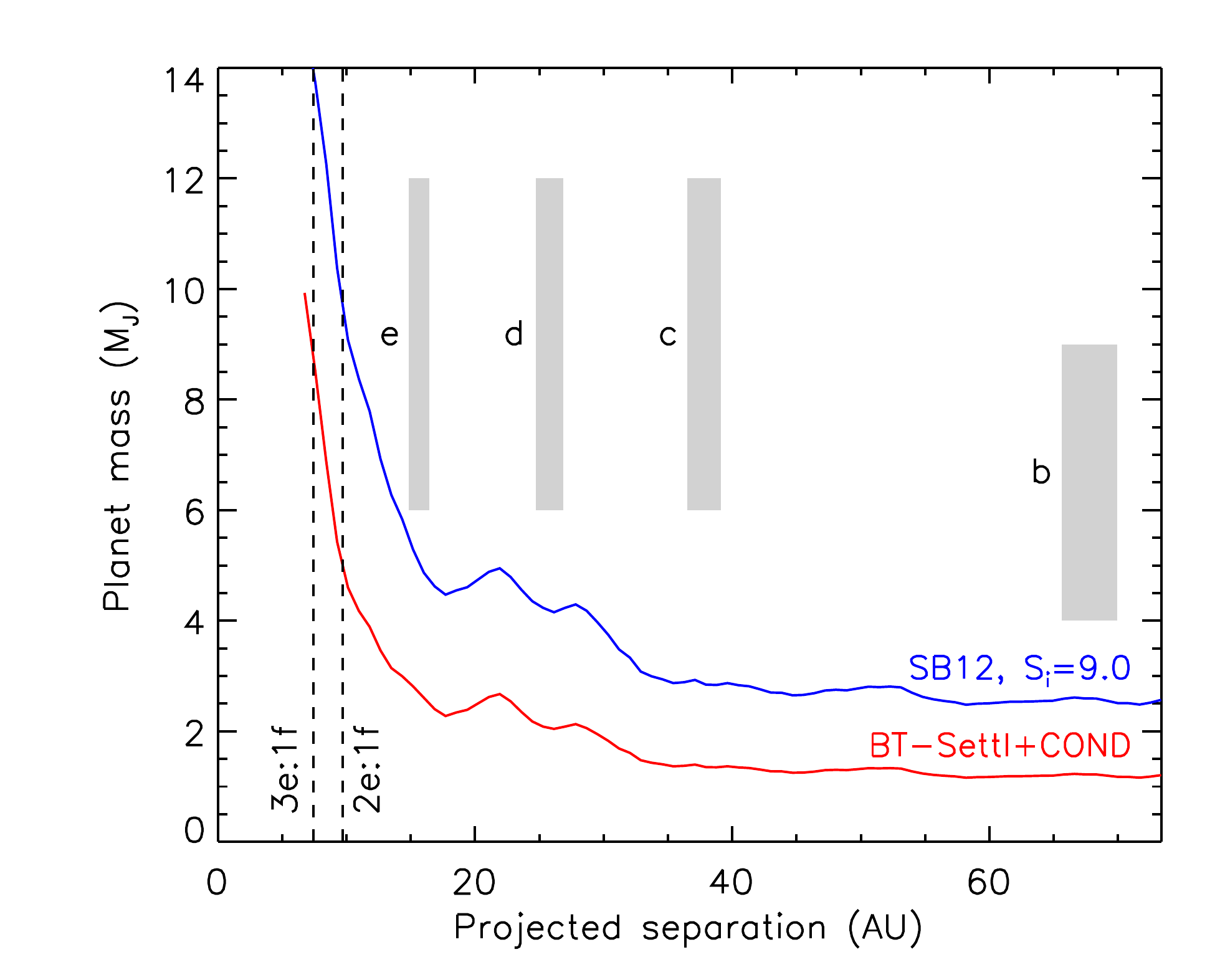}
      \caption{{5-$\sigma$ detection limits in $L^{\prime}$ band for the LMIRCam data of HR~8799 expressed in contrast (\textit{left}) and in planet mass (\textit{right}), with the four planets shown (filled circles {in the \textit{left} panel and grey areas in the \textit{right} panel}). The positions of a putative fifth planet in resonance {3:1 and 2:1} with planet e according to the models of \citet{Gozdziewski2014} are also indicated (assuming a face-on and circular orbit). For the \textit{right} plot, the detection limits are derived for an age of 30~Myr from the ``hot-start'' COND model \citep{Baraffe2003} coupled to the atmosphere model BT-Settl \citep{Allard2012} and the ``warm-start'' evolutionary+atmosphere model of \citet{Spiegel2012} for an initial entropy of 9~$k_{\rm{B}}$/baryon (see text). The range of masses for HR~8799 bcde is delimited by the predictions of the two evolutionary models assumed for deriving the detection limits (lower limit: BT-Settl+COND; upper limit: SB12 with initial entropy of 9~$k_{\rm{B}}$/baryon). The range of physical separations for the planets is computed using the errors on the planet separation and the star distance.}
                    }
         \label{fig:contrastcurve}
   \end{figure*}

We did not acquire unsaturated images of HR~8799 during the observations, hence we derive the relative photometry of planets bde with respect to planet c and use the absolute photometry of HR~8799 c reported in \citet{Marois2008c} and its uncertainty (0.09~mag) to determine the absolute photometry and the associated error bars. {We combine two sets of photometric measurements to derive the values reported in Table~\ref{tab:photometry}. The statistical error on the individual measurements includes the fitting error and the stellar flux variations during the observation. We estimate the latter errors to be 1.1\% using the photometric variations of an unsaturated optical ghost in the frames.} Given the error bars, our values are in agreement with the values in the literature and achieve similar accuracies \citep{Marois2008c, Marois2010b, Currie2014}. For these reasons, we do not attempt to carry out atmospheric modeling \citep{Skemer2012, Skemer2014b} using these new photometric data.

\section{Constraints on the properties of a fifth planet}
\label{sec:detlims}

We now discuss the constraints on the separation and mass of a putative fifth planet closer-in, ``HR~8799 f'', based on the data presented in this paper and the theoretical predictions in \citet{Gozdziewski2014}. We show the 5-$\sigma$ detection contrast curve measured in the reduced image (after removing the signals of the four planets) and corrected for the self-subtraction of off-axis point sources \citep[][and Sect.~\ref{sec:observations}]{Skemer2012} in the left panel of Fig.~\ref{fig:contrastcurve}, as well as the four known planets. The locations of the resonances {3:1 and 2:1} with planet e according to the models of \citet{Gozdziewski2014} are also indicated, assuming for simplicity a face-on and circular orbit for planet ``f''. For resonance 3e:1f, \citet{Gozdziewski2014} predict a separation and a mass range for ``HR~8799 f'' of $\sim$7.4~AU and $\sim$2--8 Jupiter masses ($M_{J}$), while for resonance 2e:1f, they derive values of $\sim$9.7~AU and $\sim$1.5--5~$M_{J}$. Given the contrast curve in the left panel of Fig.~\ref{fig:contrastcurve}, we are able to exclude a fifth planet slightly brighter/more massive than planet b at the location of 2e:1f resonance and a fifth planet about twice as bright as planets cde at the location of 3e:1f resonance.

We show in the right panel of Fig.~\ref{fig:contrastcurve} the detection limits in planet mass. We assume an age for HR~8799 of 30~Myr \citep{Marois2010b, Baines2012}. The masses of HR~8799 bcde and the detection limits are derived from two models: the ``hot-start'' model COND \citep{Baraffe2003} coupled with the atmosphere model BT-Settl \citep{Allard2012} and the ``warm-start'' evolutionary+atmosphere model of \citet{Spiegel2012} corresponding to an initial entropy of 9~$k_{\rm{B}}$/baryon and a cloudy atmosphere of one solar metallicity\footnote{We note that the atmospheric properties (presence/absence of clouds, metallicity) also affect an object's luminosity, but the effect is less significant with respect to the choice of initial conditions (initial entropy).}. The value for the initial entropy is selected based on the work of \citet{Marleau2014}, assuming an upper limit for the planet masses of 12~$M_{\rm{J}}$ according to the dynamical analysis of \citet{Gozdziewski2014}\footnote{{The entropy value of 9~$k_{\rm{B}}$/baryon refers to the predictions of \citet{Spiegel2012}. To obtain this value, we decreased the value predicted by \citet{Marleau2014} by an offset of 0.45~$k_{\rm{B}}$/baryon. This offset stems from different assumptions for the equation of state between the predictions of \citet{Spiegel2012} and \citet{Marleau2014} (see \citet{Marleau2014} for details).}}. {The selected value for the initial entropy is thus the approximate minimum value allowed for the formation of the planets independently from the details of the formation scenario.} {We cannot reject a ``warm-start'' model for the formation and evolution of the HR~8799 planets given the dynamical mass constraints.} The constraints on a planet mass for a given initial entropy still {depend} on the assumption on the star age. Thus, the detection limits in the right panel of Fig.~\ref{fig:contrastcurve} define the range of possible values, assuming HR~8799 is a 30-Myr star. The range will be shifted towards larger masses when assuming older ages. However, constraints from the stellar properties \citep{Marois2008c, Baines2012}, the mass of the disk \citep{Su2009}, and models of dynamical stability of the four planets \citep{Esposito2013, Gozdziewski2014} suggest that the system is young. {Other hints for small radii/masses of the planets come from the fitting of their spectral energy distribution \citep[e.g.,][]{Marois2008c, Bowler2010, Currie2011, Barman2011a, Madhusudhan2011, Galicher2011, Ingraham2014}, although there are uncertainties on the atmospheric composition and the physical processes governing their atmospheres assumed for the models (e.g., solar/non-solar metallicity, physics of the clouds, non-equilibrium chemistry).}

Directly South of the star {at $\sim$0.2$''$}, we find an excess at the $\sim$3--4~$\sigma$ level (see Fig.~\ref{fig:hr8799image}), which is below our planet detection threshold. {The excess is detected at higher signal-to-noise ratio for one sequence recorded at one nod. Thus, it could be a speckle correlated between the nods. The excess} is not seen in other LBTI datasets from the same run, although these datasets were taken in worse conditions. As noted by \citet{Mawet2014}, high standard deviation residuals become increasingly common at small separations, so we note the excess, but categorize it as most likely a PSF residual. We do not detect any excess at the location of the residual reported in \citet{Currie2014}.

   \begin{figure*}[t]
   \centering
   \includegraphics[trim = 1mm 3mm 1.5mm 1mm, clip,width=.95\textwidth]{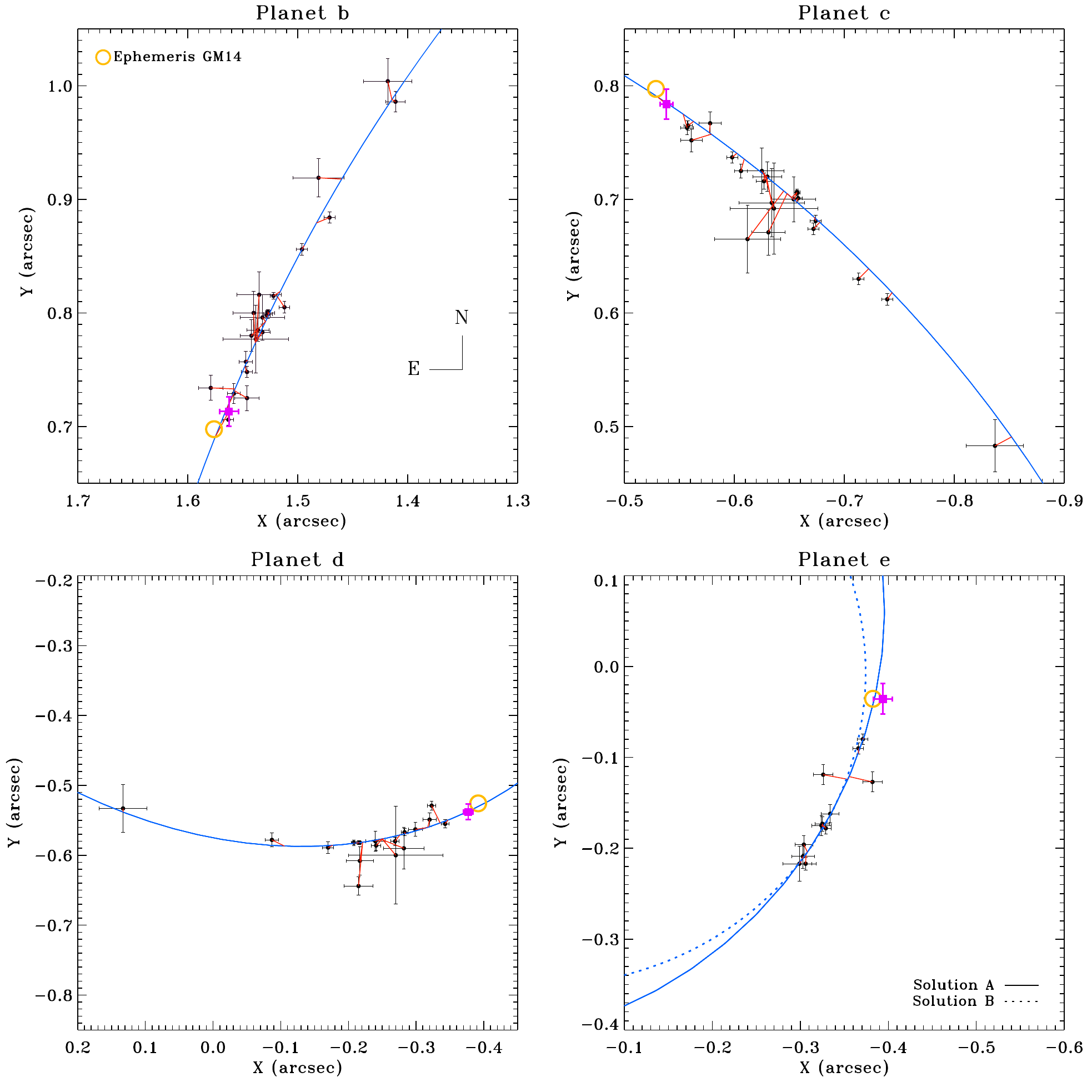}
      \caption{{Relative astrometry of the HR~8799 planets for all the measurements available in the literature (small filled circles) and the measurements derived in this work (purple filled squares)}. {The solid blue lines represent the orbital solution labeled ``A'' in Table~\ref{tab:orbits}.} Red lines connect the predicted and observed positions for all the data points. {The dotted line in the panel for HR~8799~e indicates the orbital solution labeled ``B'' in Table~\ref{tab:orbits}. The orange circles} indicate the ephemeris predicted in \citet{Gozdziewski2014}.
                    }
         \label{fig:hr8799plots}
   \end{figure*}

\section{Orbital architecture}
\label{sec:orbitarchitect}

{Mean motion resonances have been proposed to maintain the dynamical stability of the compact four-giant planet system of HR~8799 on timescales of at least the star lifetime \citep[$>$30~Myr,][]{Marois2010b}. The architecture of the system is still not completely understood and two families of orbital solutions have been investigated in the literature: non-circular and coplanar orbits \citep{Gozdziewski2009, Fabrycky2010, Soummer2011, Gozdziewski2014} and circular and non-coplanar orbits \citep{Esposito2013}. Recently, \citet{Gozdziewski2014} built a long-term stable model of the system, with all planets on quasi-circular coplanar orbits and likely involved in 8b:4c:2d:1e mean motion resonance. We first test the predictions of \citet{Gozdziewski2014} with respect to the astrometric measurements presented in this paper. We gather all the relative astrometric measurements available in the literature \citep{Marois2008c, Lafreniere2009, Fukagawa2009, Metchev2009, Hinz2010, Marois2010b, Currie2011, Bergfors2011, Galicher2011, Soummer2011, Currie2012, Esposito2013, Currie2014, Pueyo2014} and include the LEECH measurements (Table~\ref{tab:astrometry}) to represent Fig.~\ref{fig:hr8799plots}. We also indicate in Fig.~\ref{fig:hr8799plots} the ephemeris predicted by \citet{Gozdziewski2014} for a four-planet system (see their Appendix B.1) at the epoch of the observations presented in this paper. The presence of a putative fifth planet interior to the known planets does not affect the ephemeris of the latter significantly. Given the error bars, our measurements are in agreement within 3~$\sigma$ with the ephemeris.}

{To complement the analysis of \citet{Gozdziewski2014}, based on astrometric data obtained before 2012, we perform new orbital fitting of the planets assuming circular and non-coplanar orbits using the more extended baseline offered by the measurements of \citet{Currie2014}, \citet{Pueyo2014}, and the data reported in this work. We use the orbital fitting procedure described in \citet{Esposito2013}, i.e. a best-fit least-squares orbital fitting of the four planets simultaneously based on the Levenberg-Marquardt algorithm (see note~\ref{note:mpfit}). We do not intend to perform an in-depth orbital analysis like the study presented in \citet{Pueyo2014}, because such an analysis would soon be refined with the availability of more recent astrometric data.}

\begin{table}[t]
\caption{{Orbital elements fitted on the astrometric data in Fig.~\ref{fig:hr8799plots}.}}
\label{tab:orbits}
\begin{center}
\begin{tabular}{l c c c}
\hline\hline
Parameter & A & B \\
\hline
$P_{\rm{b}}$ (yr) & 456.12\,$\pm$\,2.81 & 456.12\,$\pm$\,2.81 \\
$i_{\rm{b}}$ ($^{\circ}$) & 18.50\,$\pm$\,1.01 & 18.50\,$\pm$\,1.01 \\
$\Omega_{\rm{b}}$ ($^{\circ}$) & 52.37\,$\pm$\,10.31 & 52.37\,$\pm$\,10.31 \\
$e_{\rm{b}}$ & -- & -- \\
$\omega_{\rm{b}}$ ($^{\circ}$) & -- & -- \\
$T0_{\rm{b}}$ (yr) & 1995.23\,$\pm$\,13.88 & 1995.23\,$\pm$\,13.88 \\
$a_{\rm{b}}$ (AU) & 67.98\,$\pm$\,0.28 & 67.98\,$\pm$\,0.28 \\
\hline
$P_{\rm{c}}$ (yr) & 228.06\,$\pm$\,1.41 & 228.06\,$\pm$\,1.41 \\
$i_{\rm{c}}$ ($^{\circ}$) & 29.01\,$\pm$\,0.47 & 29.01\,$\pm$\,0.47 \\
$\Omega_{\rm{c}}$ ($^{\circ}$) & 61.83\,$\pm$\,1.90 & 61.83\,$\pm$\,1.90 \\
$e_{\rm{c}}$ & -- & -- \\
$\omega_{\rm{c}}$ ($^{\circ}$) & -- & -- \\
$T0_{\rm{c}}$ (yr) & 1845.87\,$\pm$\,1.56 & 1845.87\,$\pm$\,1.56 \\
$a_{\rm{c}}$ (AU) & 42.82\,$\pm$\,0.18 & 42.82\,$\pm$\,0.18 \\
\hline
$P_{\rm{d}}$ (yr) & 114.03\,$\pm$\,0.70 & 114.03\,$\pm$\,0.70 \\
$i_{\rm{d}}$ ($^{\circ}$) & 37.33\,$\pm$\,0.48 & 37.33\,$\pm$\,0.48 \\
$\Omega_{\rm{d}}$ ($^{\circ}$) & 57.94\,$\pm$\,1.30 & 57.94\,$\pm$\,1.30 \\
$e_{\rm{d}}$ & -- & -- \\
$\omega_{\rm{d}}$ ($^{\circ}$) & -- & -- \\
$T0_{\rm{d}}$ (yr) & 1965.68\,$\pm$\,0.28 & 1965.68\,$\pm$\,0.28 \\
$a_{\rm{d}}$ (AU) & 26.98\,$\pm$\,0.11 & 26.98\,$\pm$\,0.11 \\
\hline
$P_{\rm{e}}$ (yr) & 57.02\,$\pm$\,0.35 & 46.24\,$\pm$\,0.38 \\
$i_{\rm{e}}$ ($^{\circ}$) & 31.16\,$\pm$\,1.61 & 20.76\,$\pm$\,4.85 \\
$\Omega_{\rm{e}}$ ($^{\circ}$) & 140.30\,$\pm$\,8.60 & 82.52\,$\pm$\,11.34 \\
$e_{\rm{e}}$ & -- & -- \\
$\omega_{\rm{e}}$ ($^{\circ}$) & -- & -- \\
$T0_{\rm{e}}$ (yr) & 1994.74\,$\pm$\,1.16 & 1990.28\,$\pm$\,1.47 \\
$a_{\rm{e}}$ (AU) & 16.99\,$\pm$\,0.07 & 14.78\,$\pm$\,0.08 \\
\hline
$\sqrt{\chi_{\mathrm{red,b}}^2}$ & 1.15 & 1.15 \\
$\sqrt{\chi_{\mathrm{red,c}}^2}$ & 1.17 & 1.17 \\
$\sqrt{\chi_{\mathrm{red,d}}^2}$ & 1.45 & 1.45 \\
$\sqrt{\chi_{\mathrm{red,e}}^2}$ & 0.92 & 1.08 \\
\hline
$M_{\mathrm{star}}$ $(M_{\odot})$ & 1.51 & 1.51 \\
\hline
\end{tabular}
\end{center}
\tablefoot{{Case A: Orbital solution setting circular orbits and 8b:4c:2d:1e mean motion resonance. Case B: Orbital solution setting circular orbits and 4b:2c:1d and 5d:2e mean motion resonances.\\
For each planet, the notations refer to the orbital period, inclination, longitude of the ascending node, eccentricity, argument of periapsis, time of periapsis passage, and semi-major axis. {The error bars on the orbital parameters are at 1~$\sigma$ (see text).} We also indicate the square root of the reduced $\chi^2$ {for the nominal orbital solution for each planet} and the mass assumed for the host star for the orbital fitting (see text). The semi-major axes are derived using Kepler's third law assuming the fitted orbital periods and that the planet masses are negligible with respect to the mass of the host star.}} \\
\end{table}

{We show in Fig.~\ref{fig:hr8799plots} the orbital fits (Table~\ref{tab:orbits}) assuming two orbital solutions proposed by \citet{Esposito2013}: A/ circular and non-coplanar orbits, with all planets involved in 8b:4c:2d:1e mean motion resonance, and B/ circular and non-coplanar orbits, assuming 4b:2c:1d and 5d:2e mean motion resonances. {\citet{Esposito2013} are not able to distinguish them with the available data, but predict that the orbits would diverge significantly within 2--3 years. We assume the same ranges of allowed values {for the orbital elements} as found by \citet{Esposito2013} using Monte Carlo simulations (see their Fig.~8). The semi-major axes are derived assuming a mass for the host star of 1.51~$M_{\odot}$, following the work of \citet{Baines2012}. {We estimate the error bars on the orbital parameters in Table~\ref{tab:orbits} based on the sensitivity of the orbital fits with the errors on the astrometric data (Fig.~\ref{fig:hr8799plots}). We generate 1000 random sets of astrometric measurements for each planet assuming gaussian distributions around the nominal astrometric measurements. We then perform the fitting of these measurements assuming the properties of each orbital solution. Except for a few solutions, all the fitted solutions produce a square root of the reduced $\chi^2$ below 2 for each planet. There are no clear minima in the distribution of the orbital parameters. The error bar on each orbital parameter is the standard deviation of the fitted values.} The astrometric measurements obtained in 2012 by \citet{Pueyo2014} and \citet{Currie2014} and in 2013 by us allow a modest increase in the time baselines for the three outer planets but a significant improvement for planet e. In particular, our measurement for planet e favors the orbital solution ``A'' (Table~\ref{tab:orbits}), i.e. an orbital architecture based on the mean motion resonance 8:4:2:1. {Nevertheless, the LEECH measurement is out of solution ``B'' by $\sim$2~$\sigma$ only, so this solution is not firmly excluded from a statistical point of view.} We note that \citet{Gozdziewski2014} find that the most likely mean motion resonance for the planets is 8:4:2:1 under the assumption of quasi-circular and coplanar orbits.}

\section{Discussion}
\label{sec:discussion}

{We discuss in this section the assumptions and the results for the planet architecture described in Sect.~\ref{sec:orbitarchitect} in the light of the recent analysis of \citet{Pueyo2014}. These authors performed a Bayesian analysis based on Markov Chain Monte Carlo techniques of the astrometric measurements of all the four planets compiled in \citet{Esposito2013} and P1640 measurements obtained in June 2012. The main motivation of their work is to complement the previous analyses, which are based on least-squares orbital fitting and/or dynamical studies and consider strong assumptions for the orbits (coplanarity, circularity, and/or mean motion resonances). These assumptions are used because of the degeneracies inherent to the determination of the six Keplerian elements of an orbit based on data covering only a small portion of the latter. However, as outlined by \citet{Pueyo2014}, Markov Chain Monte Carlo techniques are sensitive to underestimated biases which are non accounted for in the error bars\footnote{Nevertheless, as noted by \citet{Pueyo2014}, these biases could be included in the Markov Chain Monte Carlo state vectors.}. Such biases can come from measurements obtained with different instruments and/or data analyses.}

{Our work belongs to the category of orbital fitting studies and is based on the assumption of circular and non-coplanar orbits and the use of Monte Carlo simulations for the determination of broad ranges for the orbital parameters. Monte Carlo methods are sensitive to the same kind of biases as Markov Chain Monte Carlo methods, hence we cannot exclude biases in our results, in particular the mean motion resonances between the planets.}

{We first focus on the relative inclinations of the planets. We consider in this paper non-coplanar orbits. Our estimates in Table~\ref{tab:orbits} do not allow to conclude that the orbit of one planet is out of the plane of the orbits of the other planets. \citet{Pueyo2014} find that planets bce would have similar inclinations}, but that planet d would have a more inclined orbit ($\sim$15--20$^{\circ}$). Looking at Fig.~\ref{fig:hr8799plots}, we note the discrepancy of the P1640 data for planet d ($\Delta$RA\,=\,$-$0.323$\pm$0.006$''$, $\Delta$Dec\,=\,$-$0.529$\pm$0.006$''$) with respect to the October 2012 data reported in \citet{Currie2014} ($\Delta$RA\,=\,$-$0.343$\pm$0.006$''$, $\Delta$Dec\,=\,$-$0.555$\pm$0.006$''$) and our October 2013 measurement (purple filled square). The P1640 measurement is likely the cause for the orbital solutions found for this planet by \citet{Pueyo2014}. {Our motivation for exploring non-coplanar orbits for the planets in \citet{Esposito2013} and this work is to complement the available studies, which assume coplanarity. It is also suggested by asteroseismology studies indicating that the star has a more edge-on inclination \citep[$\geq$40$^{\circ}$,][]{Wright2011} with respect to the planes of the planet orbits and of the disk \citep{Su2009}. A recent analysis of the MOST photometric data of the star \citep{Sodor2014} does not identify clear evidences of rotational splitting of the modes, which would imply a stellar inclination similar to those of the planes of the planet orbits and of the disk. Thus, we cannot draw sound conclusions for a misalignment/alignment of the star equator with respect to the plane of the planet orbits. {Nevertheless, we note that under the extreme assumption of circular orbits, we find that HR~8799 bc would have orbits roughly coplanar with the disk of the host star \citep[$i$\,=\,26\,$\pm$\,3$^{\circ}$ and $\Omega$\,=\,62\,$\pm$\,3$^{\circ}$,][]{BMatthews2014}. HR~8799~d might have an orbit slightly inclined ($\sim$8--11$^{\circ}$) with respect to the disk plane, although our study does not allow to conclude on this point. According to the orbital solution favored in our analysis (solution ``A'', Table~\ref{tab:orbits}), HR~8799~e might have an orbit significantly misaligned with respect to the position angle of the disk (the difference between the longitudes of ascending node is $\sim$75--78$^{\circ}$).}

{We now test the hypothesis of circular orbits for all planets that we made in our analysis (Sect.~\ref{sec:orbitarchitect}).} Previous analyses indicate that planets bc would have roughly circular orbits, while planet d would have a moderately-eccentric orbit \citep[$\lesssim$0.15,][]{Fabrycky2010, Bergfors2011, Soummer2011, Gozdziewski2014}. For planet e, studies suggest an eccentric orbit \citep[$\lesssim$0.15,][]{Sudol2012, Gozdziewski2014}. \citet{Pueyo2014} find roughly circular orbits for planets bc, but a very eccentric ($\leq$0.3 at 1~$\sigma$) orbit for planet d. As abovementioned, the P1640 data for planet d appears discrepant with respect to the 2012 measurement by \citet{Currie2014} and probably biases their results for this planet. {We perform several orbital fitting tests assuming non-circular and coplanar orbits for all planets assuming the values of inclination ($i$\,$\sim$\,26$^{\circ}$) and longitude of ascending node ($\Omega$\,$\sim$\,62$^{\circ}$) derived in \citet{BMatthews2014} and find moderate eccentricities for planet d ($\lesssim$0.15), in agreement with the previous analyses abovementioned}. For planet e, the time baseline used by \citet{Pueyo2014} is limited so prevents them {from placing} strong constraints. Our tests suggest an eccentricity similar to that of planet d or higher (up to $\sim$0.25) depending on the assumed mean motion resonance between planets d and e.

We finally discuss the mean motion resonances. {Given the broad ranges} of orbital parameters indicated by the Monte Carlo simulations performed in \citet{Esposito2013} assuming circular orbits (see their Fig.~8), we find that the hypothesis of {4:2:1} mean motion resonance for the three outer planets produces the best-fitting solution to the data. This result is in agreement with previous analyses \citep{Fabrycky2010, Gozdziewski2009, Reidemeister2009, Soummer2011, Gozdziewski2014}. {For planets de, our analysis favors a period ratio of 2:1 over 5:2. The analysis described in \citet{Pueyo2014} rules out a {2:1} mean motion resonance for the two outermost planets and favors a period ratio between {3:1 and 5:2}. They find that a 2:1 period ratio for planets cd is still favored. Finally, \citet{Pueyo2014} suggest that the most likely period ratio for planets de would be {3:2}, but cannot rule out a value of {2:1}. We test the hypothesis of {3:2} period ratio for planets de assuming circular and non-coplanar orbits and find that it cannot be excluded given our dataset, although it would induce an orbit for planet e strongly misaligned ($i$\,$\sim$\,45$^{\circ}$ and $\Omega$\,$\sim$\,149$^{\circ}$) with respect to the orbit of planet d ($i$\,$\sim$\,37$^{\circ}$ and $\Omega$\,$\sim$\,$+$58$^{\circ}$, Table~\ref{tab:orbits}). The inclinations of planets de and the longitude of ascending node of planet e might be diminished towards values closer to those of planets bc when relaxing the constraint of circular orbits.}

\section{Conclusions}

We presented in this paper the first reliable astrometric analysis of an exoplanetary system, HR~8799, observed in the context of LEECH, the high-contrast imaging survey to search for and characterize young and adolescent exoplanets in $L^{\prime}$ band using LBTI/LMIRCam. To achieve this, we first performed the astrometric calibration of the LMIRCam camera using the $\Theta^1$ Ori C field. We characterized the distortion of this new instrument and refined the estimates of platescale and {true north} with respect to previous works, which did not account for the distortion effects. Applying the results of this analysis to HR~8799 data obtained in October 2013, we were able to further constrain the orbit of HR~8799 e, based on the previous analysis described in \citet{Esposito2013}, who used data covering only a 2.3-yr baseline between mid 2009 and late 2011. In particular, the new measurements favor an orbital model with all planets in multiple double Laplace resonance, assuming circular and non-coplanar orbits. Our results are in agreement with the recent models of \citet{Gozdziewski2014}, who assume quasi-coplanar and eccentric orbits. Astrometric monitoring in the coming years with the current and new-generation high-contrast imaging instruments will further improve the time baseline and maybe confirm these conclusions.

We also presented {deep constraints on a putative fifth planet} interior to the known planets. We are able to rule out a planet slightly brighter/more massive than HR~8799 b at the location of the {2:1} mean motion resonance with HR~8799 e and a planet twice as bright as HR~8799 cde at the location of the {3:1} mean motion resonance with HR~8799 e. However, \citet{Gozdziewski2014} predict that this fifth planet would be less massive/luminous than planet b in the first case and would have a mass inferior to 8~$M_{\rm{J}}$ in the second case. The corresponding physical separations ($\sim$7--9~AU) translate into angular separations of $\sim$0.20--0.25$''$. Detecting a faint planet at these very short separations in the coming years could be challenging, even with the new generation of high-contrast imaging instruments on 8--10-m telescopes (SPHERE, GPI, SCExAO+CHARIS).

One of the most significant errors in current astrometric measurements {from high-contrast imaging is the estimation of the star center in saturated/coronagraphic images. This is also the case for our LBTI/LMIRCam measurements, because of the saturation of the star. This error could be reduced using the adaptive secondary mirror to create several faint replica of the star at a given separation in the field of view in order to accurately derive the star center \citep{Marois2006c, Sivaramakrishnan2006}. The use of star replica for estimating the center of coronagraphic/saturated images is employed in the new high-contrast imaging instruments SPHERE \citep[][Mesa et al., A\&A, subm.]{Beuzit2012, Zurlo2014} and GPI \citep{Macintosh2014}.} On-sky tests are foreseen in the coming months to implement this technique for the LEECH observations.

{Finally, for astrometric analyses based on data with high signal-to-noise ratios, we outline that if the error budget related to the instrument (uncertainties on platescale, {true north}, and distortion) and to the estimation of the star center are reduced to levels of the order of 1--2~mas, the errors on the stellar positions in standard astrometric fields are no longer negligible.} {The main source of the latter errors are the differential proper motions of the stars between the epoch of the reference data (typically \textit{Hubble Space Telescope} data) and the epoch of the science observations to be calibrated.} It would be very useful to define a common set of reliable astrometric calibrators to be used for the on-going or upcoming high-contrast imaging surveys (P1640, LMIRCam, GPI, SPHERE, MagAO), to improve the accuracy of analyses based on combinations of datasets. We note that the \textit{Gaia} all-sky survey will contribute significantly in providing good astrometric calibrators in the coming years.

\begin{acknowledgements}
      {We thank the referee for a detailed and constructive report which helped to improve the manuscript.} We thank Thayne Currie and Gabriel-Dominique Marleau for useful comments. A.-L.M. thanks Arthur Vigan for helping on distortion correction and Dimitri Mawet for discussions on contrast estimation at small separations. A.-L.M., S.D., R.G., R.U.C., and D.M. acknowledge support from the ``Progetti Premiali'' funding scheme of the Italian Ministry of Education, University, and Research. Support for A.J.S. was provided by the National Aeronautics and Space Administration through Hubble Fellowship grant HST-HF2-51349 awarded by the Space Telescope Science Institute, which is operated by the Association of Universities for Research in Astronomy, Inc., for NASA, under contract NAS 5-26555. E.B. was supported by the Swiss National Science Foundation (SNSF). The research of J.E.S was supported in part by an appointment to the NASA Postdoctoral Program at NASA Ames Research Center, administered by Oak Ridge Associated Universities through a contract with NASA. LEECH is funded by the NASA Origins of Solar Systems Program, grant NNX13AJ17G. The Large Binocular Telescope Interferometer is funded by NASA as part of its Exoplanet Exploration program. LMIRCam is funded by the National Science Foundation through grant NSF AST-0705296.
\end{acknowledgements}

\bibliographystyle{aa}
\bibliography{biblio}

%-------------------------------------------------------------------

\begin{appendix}

\section{Distortion correction}
\label{sec:appendixdistortion}

\begin{table*}[!t]
\caption{Coefficients of the distortion solution for the LMIRCam camera.}
\label{tab:distortion}
\begin{center}
\begin{tabular}{l c c c c c c c c c c}
\hline\hline
$a_{i}$ & $-2.148$ & 1.011 & 5.814e-3 & $-$2.116e-5 & $-$2.383e-5 & $-$7.640e-6 & 1.298e-8 & 5.312e-8 & 2.846e-8 & 2.536e-9 \\ 
\hline
$b_{i}$ & 9.272 & $-$1.362e-2 & 0.988 & 1.131e-5 & $-$3.953e-5 & 4.351e-6 & 1.628e-9 & 6.713e-8 & 8.120e-8 & 9.345e-9 \\ 
\hline
\end{tabular}
\end{center}
\end{table*}

A new distortion grid was installed in LMIRCam late July 2014. The purpose of this grid is to measure the distortion effects of the camera, which are expected to be much more significant than the distortion induced by the telescope and the light combiner. The grid is formed of holes of diameter 381~$\muup$m regularly spaced by 838~$\muup$m ($<$6\% accuracy) following a square pattern. The sampling of the grid spots in the images is $\sim$48~pixels. We give in Table~\ref{tab:distortion} the coefficients of the polynomial laws used to correct for the measured distortion effects. The solution is accurate to $\sim$0.1\% on the separation at 1~$\sigma$. We expect the distortion to be roughly constant with time, depending on the optical properties, so we use this distortion calibration for the HR~8799 data taken in October 2013. In the future, the distortion calibration will be regularly performed to monitor passive temporal variations due to, e.g., temperature variations. Assuming the position of the image center is ($x_0$,$y_0$), the distortion-corrected coordinates ($x$,$y$) are related to the raw coordinates ($x^{\prime}$,$y^{\prime}$) by:

\begin{align}
x^\prime=&~a_0\,+\,a_1(x-x_0)\,+\,a_3(x-x_0)^2\,+\,a_6(x-x_0)^3 \nonumber \\
&+\,a_2(y-y_0)\,+\,a_4(x-x_0)(y-y_0)\,+\,a_7(x-x_0)^2(y-y_0) \nonumber \\
&+\,a_5(y-y_0)^2\,+\,a_8(x-x_0)(y-y_0)^2\,+\,a_9(y-y_0)^3,
\end{align}
\begin{align}
y^\prime=&~b_0\,+\,b_1(x-x_0)\,+\,b_3(x-x_0)^2\,+\,b_6(x-x_0)^3 \nonumber \\
&+\,b_2(y-y_0)\,+\,b_4(x-x_0)(y-y_0)\,+\,b_7(x-x_0)^2(y-y_0) \nonumber \\
&+\,b_5(y-y_0)^2\,+\,b_8(x-x_0)(y-y_0)^2\,+\,b_9(y-y_0)^3.
\end{align}

\end{appendix}

\end{document}